\documentclass[10pt]{article}

\usepackage{amsmath}
\usepackage{amssymb}
\usepackage{amsthm}
\usepackage{arydshln}
\usepackage{graphicx}
\usepackage{newclude}

\makeatletter
\newtheorem*{rep@theorem}{\rep@title}
\newcommand{\newreptheorem}[2]{%
\newenvironment{rep#1}[1]{%
\def\rep@title{#2 \ref{##1}}%
\begin{rep@theorem}}%
{\end{rep@theorem}}}
\makeatother

\newtheorem{theorem}{Theorem}[section]
\newtheorem{lemma}[theorem]{Lemma}

\newtheorem{corollary}[theorem]{Corollary}
\newtheorem*{definition}{Definition}

\newreptheorem{theorem}{Theorem}
\newreptheorem{lemma}{Lemma}
\newreptheorem{proposition}{Proposition}
\newreptheorem{corollary}{Corollary}

\newtheoremstyle{algorithm}
  {\topsep}                   
  {\topsep}                   
  {\normalfont}               
  {0pt}                       
  {\bfseries}                 
  {: }                        
  {5 pt plus 1 pt minus 1 pt} 
  {}                          
\theoremstyle{algorithm}
  \newtheorem*{correctness}{Correctness}
  \newtheorem*{runningtime}{Running Time}
  \newtheorem*{memoryusage}{Memory Usage}

\begin{document}

\title{Space-Efficient Las Vegas Algorithms for K-SUM (Preliminary Version)}
\author{Joshua Wang\\
        Stanford University,\\
        \texttt{joshua.wang@cs.stanford.edu}}
\date{\today}
\maketitle

\begin{abstract}
  Using hashing techniques, this paper develops a family of space-efficient Las
  Vegas randomized algorithms for $k$-SUM problems. This family includes an
  algorithm that can solve 3-SUM in $O(n^2)$ time and $O(\sqrt{n})$ space. It
  also establishes a new time-space upper bound for SUBSET-SUM, which can be
  solved by a Las Vegas algorithm in $O^*(2^{(1-\sqrt{\frac{8}{9}\beta})n})$
  time and $O^*(2^{\beta n})$ space, for any $\beta \in [0, \frac{9}{32}]$.
\end{abstract}

\section{Introduction}

The $k$-SUM problem on $n$ numbers can be formulated as follows: Given $k$ sets
$S_1$, $S_2$, ..., $S_k$ with $n$ integers each and a target $t$, find $a_1,
a_2, \ldots, a_k$ such that for all $i$, $a_i \in S_i$ and $\sum_{i=1}^k a_i =
t$. Note that one common variant of the problem has only a single set $S$ from
which all elements in the solution are chosen from, but the two are easily
reducible to each other. The $k$-SUM problem can be trivially solved in $O(n^k)$
arithmetic operations by trying all possibilities, and a more sophisticated
solution runs in $O(n^{\lceil k/2 \rceil} \log n)$ time. How
much faster can it be solved? This turns out to be a fundamental question, as
the complexity of $k$-SUM is related to the complexity of a number of other
problems.

Gajentaan and Overmars \cite{Gajentaan:2012:COP:2109239.2109580} classified
many problems from computational geometry as ``3SUM-hard'' (i.e. there exists a
$o(n^2)$ reduction from 3-SUM to the problem in question) in order to indirectly
demonstrate their difficulty. Finding a subquadratic algorithm for any problem
in this class of problems would immediately produce a subquadratic algorithm for
3-SUM. One example of such a problem is 3-POINTS-ON-LINE: Given a set of points
in the plane, are there three collinear points? To reduce 3-SUM to this problem,
map each $x \in S$ (using the single-set variation of 3-SUM) to the point
$(x, x^3)$, with the idea that $a_1 + a_2 + a_3 = 0$ if and only if the points
$(a_1, a_1^3)$, $(a_2, a_2^3)$, and $(a_3, a_3^3)$ are collinear.

$k$-SUM is also fundamentally connected to several NP-hard problems. Patrascu
and Williams \cite{Patrascu:2010:PFS:1873601.1873687} show that solving $k$-SUM
over $n$ numbers in $O(n^{o(k)})$ time would imply that 3-SAT with $n$ variables
can be solved in $O(2^{o(n)})$ time. Schroeppel and Shamir
\cite{Schroeppel:1979:TTT:1382433.1382632} have shown how the SUBSET-SUM problem
can be reduced to an (exponentially-sized) $k$-SUM problem. Therefore, more
efficient $k$-SUM algorithms can be used to derive faster SUBSET-SUM algorithms.
The SUBSET-SUM problem on $n$ numbers can be formulated as follows: Given a set
$S$ of $n$ integers and a target $t$, find a subset $S' \subseteq S$ such that
$\sum_{a \in S'} a = t$. They then provide a space-efficient 4-SUM algorithm to
yield a time and space efficient SUBSET-SUM algorithm. Schroeppel and Shamir
also established a time-space tradeoff theorem for SUBSET-SUM algorithms that
allowed them to provide a time/space upper bound of $T \cdot S^2 = O^*(2^n)$
given that $T \ge O^*(2^{n/2})$. This paper will prove a parallel tradeoff result
for $k$-SUM algorithms, and then use their SUBSET-SUM to $k$-SUM reduction to
find an improved time-space upper bound for SUBSET-SUM.

\subsection*{Our Results}

The best known algorithm for 3-SUM takes $O(n^2)$ time, but also requires
$O(n)$ space (to hold a sorted array of numbers). Can we use significantly less
space and obtain the same running time? This paper also investigates the
time-space tradeoffs for the general $k$-SUM problem. Given some fixed time
budget $S$, we wish to solve $k$-SUM in time $T$ and space $S$ where $T$ is
minimized.

We use hashing techniques to lower the space requirement for 3-SUM:

\begin{theorem}
\label{3-SUM-better}
  3-SUM on $n$ numbers can be solved by a Las Vegas algorithm\footnote{Recall
  that algorithms are Las Vegas randomized if they always give correct results,
  but may take additional running time depending on the random numbers
  generated (but not depending on the choice of input).} in time $O(n^2)$ and
  space $O(\sqrt{n})$.
\end{theorem}

These techniques also help lower the space requirements for the general
$k$-SUM problem on $n$ numbers, albeit at the cost of some running time
increase.

\begin{theorem}
\label{las-vegas-final}
  Let $\delta \le 1$. We will not define $f(x)$ here, but it is a function from
  $\mathbb{Z}^+ \to \mathbb{Z}^+$, and $f(x) \le x - \sqrt{x} + 1$. $k$-SUM on
  $n$ numbers can be solved in 
  $\tilde{O}(n^{k-\delta(k-1)} + n^{k-\delta(k-1) + (\delta f(k)-1)})$ time and
  $O(n^{\delta})$ space by a Las Vegas algorithm.
\end{theorem}

The bound on $f(x)$ implies the following corollary when we let $\delta=1$:

\begin{corollary}
\label{sqrt-linear}
  $k$-SUM on $n$ numbers can be solved in $\tilde{O}(n^{k-\sqrt{k}+1})$ time and
  $O(n)$ space by a Las Vegas algorithm.
\end{corollary}

Here are a few sample values of $f$: $f(3)=2$, $f(4)=2$, $f(10)=7$, and
$f(100)=90$. Substituting these values into Theorem~\ref{las-vegas-final} yields:

\begin{corollary}
  Let $\delta \le 1$. Then 3-SUM on $n$ numbers can be solved in 
  $\tilde{O}(n^{3-2\delta} + n^{2})$ time and $O(n^{\delta})$ space by a Las
  Vegas algorithm.
\end{corollary}
\begin{corollary}
  Let $\delta \le 1$. Then 4-SUM on $n$ numbers can be solved in 
  $\tilde{O}(n^{4-3\delta} + n^{3-\delta})$ time and $O(n^{\delta})$ space by a
  Las Vegas algorithm.
\end{corollary}
\begin{corollary}
  Let $\delta \le 1$. Then 10-SUM on $n$ numbers can be solved in 
  $\tilde{O}(n^{10-9\delta} + n^{9-2\delta})$ time and $O(n^{\delta})$ space by
  a Las Vegas algorithm.
\end{corollary}
\begin{corollary}
  Let $\delta \le 1$. Then 100-SUM on $n$ numbers can be solved in 
  $\tilde{O}(n^{100-99\delta} + n^{99-9\delta})$ time and $O(n^{\delta})$ space
  by a Las Vegas algorithm.
\end{corollary}

These space-efficient algorithms also imply new time/space upper-bounds for
SUBSET-SUM:

\begin{theorem}
\label{approximation-curve}
  There is a Las Vegas algorithm for SUBSET-SUM on $n$ numbers that runs in
  $O^*(2^{(1-\sqrt{\frac{8}{9}\beta})n})$ time and $O^*(2^{\beta n})$
  space, for $\beta \in [0, \frac{9}{32}]$.
\end{theorem}

This improves the tradeoff of Schroeppel and Shamir when $S$ is sufficiently
small. For example, when $S = O^*(2^{0.1n})$, Schroeppel and Shamir obtain \\
$T = O^*(2^{0.8n})$ while we obtain $T = O^*(2^{0.702n})$.

\section{Preliminaries}

This section covers notation, basic $k$-SUM algorithms, and hashing.

\subsection{Notation}
Suppression of polylogarithmic factors from polynomial functions is indicated
with $\tilde{O}$. Suppression of polynomial factors from exponential functions
is indicated with $O^*$.

The following definition is also useful for discussing merging the sets of
$k$-SUM problems:
\begin{definition}
\label{minkowski-sum}
  When $S$ and $T$ are sets, the set $S + T$, also called the Minkowski sum of
  $S$ and $T$, is defined as $\{s + t \mid s \in S, t \in T\}$.
\end{definition}

\subsection{Basic $k$-SUM Algorithms}
We present several standard algorithms for $k$-SUM on $n$ numbers for $k \le 4$.
All of them are based around the following solution to 2-SUM that requires
the input sets to be sorted:

\begin{lemma}
\label{2-table}
  Given a 2-SUM problem on $n$ numbers where the elements of $S_1$ can be
  accessed in nondecreasing order and the elements of $S_2$ can be accessed in
  nonincreasing order, where $T(n)$ is the time to access the next element of
  either $S_1$ or $S_2$, a solution can be found in $O(n \cdot T(n))$ time and
  $O(1)$ space.
\end{lemma}

\begin{proof}
  Let $s_1$ denote an element of $S_1$ and $s_2$ denote an element of $S_2$.
  Begin by setting $s_1$ to the smallest element of $S_1$ and setting $s_2$
  to the largest element of $S_2$. If $s_1 + s_2 = t$, then $s_1$ and $s_2$ form
  a solution; return it. If $s_1 + s_2 < t$, then advance $s_1$ to the next
  element of $S_1$. Otherwise, if $s_1 + s_2 > t$, then advance $s_2$ to the
  next element of $S_2$. Repeat this process until a solution is found or one of
  the sets is empty, in which case there is no solution.

  \begin{correctness}
    Notice the algorithm processes elements of $S_1$ from smallest to largest
    elements of $S_2$ from largest to smallest. $s_1$ only advances when it
    could not sum to $t$ with any element left to be considered in $S_2$. This
    occurs because $s_1 + s_2 < t$ implies that the sum of $s_1$ with any
    element left in $S_2$ is strictly less than $t$. Similarly, $s_2$ only advances
    when it could not sum to $t$ wit hany element left to be considered in $S_1$.

    If the algorithm exhausts either set $S_1$ or $S_2$, then that set has
    no elements that could appear in a solution. Hence, there are no solutions.
  \end{correctness}

  \begin{runningtime}
    Each comparison with $t$ and element access removes one element to consider
    from either $S_1$ or $S_2$, so the algorithm requires $O(n \cdot T(n))$ time
    at most.
  \end{runningtime}

  \begin{memoryusage}
    This algorithm only requires space to store counters and compute sums, which
    does not depend on $n$.
  \end{memoryusage}

  This completes the proof.
\end{proof}

The algorithms for 2-SUM, 3-SUM, and 4-SUM are just reductions to the
constrained 2-SUM problem required by Lemma~\ref{2-table}:

\begin{theorem}
\label{basic-2-sum}
  2-SUM on $n$ numbers can be solved in $O(n \log n)$ time and $O(n)$ space.
\end{theorem}

\begin{proof}
  Sort the elements of $S_1$ and $S_2$ into arrays, and run the algorithm from
  Lemma~\ref{2-table}. Sorting requires $O(n \log n)$ time and $O(n)$ space,
  and note that element access can be done in constant time.
\end{proof}

\begin{theorem}
\label{basic-3-sum}
  3-SUM on $n$ numbers can be solved in $O(n^2)$ time and $O(n)$ space.
\end{theorem}

\begin{proof}
  Sort the elements of $S_1$ and $S_2$ into arrays. For each element
  $s_3 \in S_3$, use the algorithm from Lemma~\ref{2-table} to search for
  $t-s_3$.

  Sorting requires $O(n \log n)$ time and $O(n)$ space. Invoking the
  algorithm from Lemma~\ref{2-table} $n$ times requires $O(n^2)$ time and
  $O(1)$ space (element access can be done in constant time).
\end{proof}

Schroeppel and Shamir\cite{Schroeppel:1979:TTT:1382433.1382632} devised the
following 4-SUM algorithm:

\begin{theorem}
\label{basic-4-sum}
  4-SUM on $n$ numbers can be solved in $O(n^2 \log n)$ time and $O(n)$
  space.
\end{theorem}

\begin{proof}
  The key data structure is a priority queue that supports inserting, deleting,
  and extracting the minimum in logarithmic time per operation and takes linear
  space (this is possible with a heap-based priority queue). One priority
  queue, $PQ_1$, processes the elements of $S_1 + S_2$ in non-decreasing order
  while another priority queue, $PQ_2$, processes the elements of $S_3 + S_4$ in
  non-increasing order.
  
  To do this for $S_1 + S_2$, first sort $S_2$ in non-decreasing order. For
  every $i=1,2,\ldots,|S_1|$, enqueue the pair $(i,1)$. The priority of any
  pair $(i, j)$ is be $S_1[i] + S_2[j]$ (the sum of the $i^{th}$ element of
  $S_1$ and the $j^{th}$ element of the sorted $S_2$, both of which are
  one-indexed). Whenever the pair $(i, j)$ is deleted, where $j < |S_2|$,
  immediately insert the pair $(i, j+1)$. Since $S_2$ is sorted in
  non-decreasing order, any pair will be inserted before the minimum priority
  in the queue is larger than the pair's priority. The elements of $S_1 + S_2$
  are therefore extracted in order of non-decreasing priority, which is to say
  in order of non-decreasing value.

  $S_3 + S_4$ is handled similarly, except $S_4$ is sorted in non-increasing
  order and the priority queue is used to extract the maximum priority element.

  These priority queues reduce the problem to the form found in
  Lemma~\ref{2-table}, but each set now has $n^2$ elements. Accessing elements
  takes $O(\log n)$ time, so the final running time is $O(n^2 \log n)$.

  The priority queues each use linear memory and only ever contain a linear
  number of elements, so the total memory usage is $O(n)$.
\end{proof}

\subsection{Hash Functions}

\begin{definition}
\label{universal}
  A family of hash functions $H = \{h : U \to [m]\}$ is said to be universal if
  for every $x, y \in U$, if $x \ne y$ then
  $Pr_{h \in H}[h(x) = h(y)] \le \frac{1}{m}$.
\end{definition}

\begin{definition}
  Given a family of hash functions $H = \{h:U \to [m]\}$, and some set
  $S \subset U$, let the bucket of $h$ with value $v$ be $h^{-1}(\{v\})$ (i.e.
  all elements with hash value $v$). Also, define $\mathcal{B}_h(x) :=
  h^{-1}(\{h(x)\})$ (the bucket of $h$ with value $h(x)$).
\end{definition}

The following universal family of hash functions, $H_1$, was first introduced
by Dietzfelbinger
\cite{Dietzfelbinger:1996:UHK:646511.695324} and applied to 3-SUM by Baran,
Demaine and Patrascu \cite{Baran:2005:SA:2145115.2145165}. It can be used on the
elements of the input sets of $k$-SUM so that only a subset of them need to be
considered at once, saving memory.

\begin{definition}
  Given a word size $w$, a hash length $s$, and an odd integer $a$, let the
  hash function $h_a: U \to [2^s]$ be defined as $h_a(x) :=
  \lfloor \frac{ax \mod 2^w}{2^{w-s}} \rfloor$. In C notation, this hash can
  be expressed as $(a*x)>>(w-s)$.
  
  Define the family of hash functions
  $H_1 := \{h_a \mid a \in [2^w], a \text{ odd}\}$.
\end{definition}

\section{Almost Linear Hashing}

This section covers certain useful properties of $H_1$, which will be used
to construct Las Vegas algorithms for $k$-SUM. Baran, Demaine, and Patrascu
\cite{Baran:2005:SA:2145115.2145165} gave the following two lemmas when
applying $H_1$ to 3-SUM:

\begin{lemma}
\label{lemma-almost-linear}
  The family of hash functions $H_1$ satisfies almost-linearity, in that for all
  $x, y \in U$, $h(x+y) \in \{h(x)\} \oplus \{h(y)\} \oplus \{0, 1\}$ ($\oplus$
  is addition modulo $2^s$).
\end{lemma}

\begin{proof}
  Multiplying by $a$ is linear, and dropping the low-order bits can only
  influence the result by $1$ due to losing the carry.
\end{proof}

This next lemma applies to any universal family of hash functions, and hence to
$H_1$ as well:

\begin{lemma}
\label{lemma-univ}
  Given any universal family of hash functions $H = \{h:U \to [m]\}$ and some
  set $S \subset U$ of size $n$, the expected number of elements $x \in S$ with
  $|\mathcal{B}_h(x)| \ge t$ is at most $\frac{2n}{t - 2m/n + 2}$.
\end{lemma}

\begin{proof}
  Pick $x \in S, y \in S \setminus \{x\}$ randomly and let
  $p_h = Pr_x[|\mathcal{B}_h(x)| \ge t]$ and $q_h = Pr_{x, y}[h(x) = h(y)]$.
  It suffices to show that $p_h \le \frac{2}{t - 2n/m + 1}$.
  
  Let $S_h = \{x \in S \mid |\mathcal{B}_h(x)| < t\}$. Note
  $|S_h| = (1 - p_h)n$. Notice that:
  \[Pr[h(x) = h(y) \mid x \not \in S_h] \ge \frac{t-1}{n}\]
  
  On the other hand, if $x \in S_h$ then also $y \in S_h$. By
  convexity of the square function, the collision probability of elements of
  $S_h$ is minimized when the same number of elements of $S_h$ hash to any value.
  In this case:
  \[|\mathcal{B}_h(x)| \ge \lfloor \frac{|S_h|}{m} \rfloor \ge
  (1 - p_h)\frac{n}{m} - 1\]
  
  Hence:
  \[Pr[h(x) = h(y) \mid x \in S_h] \ge \frac{(1 - p_h)n/m - 2}{n}\]

  Combining yields the following:
  \begin{align*}
    q_h &\ge p_h \frac{t-1}{n} + (1 - p_h)\frac{(1 - p_h)n/m - 2}{n} \\
        &\ge \frac{1}{n}(p_h(t-1) + (1 - 2p_h)\frac{n}{m} - 2(1 - p_h)) \\
        &\ge \frac{1}{n}(p_h(t - 2\frac{n}{m} + 2) + \frac{n}{m} - 2)
  \end{align*}

  By universality, $E[q_h] \le \frac{1}{m}$. The above inequality simplifies
  into:
  \begin{align*}
    p_h(t - 2\frac{n}{m} + 2) + \frac{n}{m} - 2 &\le \frac{n}{m} \\
                                            p_h &\le \frac{2}{t - 2n/m + 2}
  \end{align*}

  This completes the proof.
\end{proof}
  
Lemma~\ref{lemma-almost-linear} guarantees that if $(k-1)$ sets have their hash
buckets fixed, any solution that uses elements from those buckets could only
have its last element in one of $k$ buckets of the last set. Hence, hashing can
be used to shrink the problem size with some limited growth in the number of
cases. It is worth noting that this hash works best on 3-SUM, since for larger
$k$ applying the hash tends to increase the running time of the algorithm. It
turns out that for large enough $m$, large buckets can be completely avoided by
simply inspecting a constant number of hashes (in expectation).

\begin{corollary}
\label{corollary-univ}
  Consider a universal family of hash functions $H = \{h:U \to [m]\}$, a set
  $S \subset U$ of size $n$, where $m \le \sqrt{n}$, and an arbitrary constant
  $c \ge 1$. Then:
  \[Pr_{h \in H}[\forall x \in S : |\mathcal{B}(x)| \le (c+2)\frac{n}{m}]
  \ge 1 - \frac{2}{c^2}\]
\end{corollary}

\begin{proof}
  Let $t = (c+2)\frac{n}{m}$. Let $b(h)$ be the number of elements $x \in S$
  with $|\mathcal{B}_h(x)| \ge t$. Applying Lemma~\ref{lemma-univ} yields that
  $E[b(h)] \le \frac{2n}{c(n/m)+2} \le \frac{2m}{c}$. Applying a Markov bound
  yields $Pr_h[ b(h) \ge cm ] \le \frac{2}{c^2}$. However, if $b(h) < cm$
  then in fact $b(h) = 0$, since $b(h)$ counts the number of elements in buckets
  of $h$ with at least $(c+2)\frac{n}{m}$ elements ($m \le \sqrt{n}$ implies
  $\frac{n}{m} \ge m$). Hence, $Pr_h[\forall x : |\mathcal{B}(x)| \le
  (c+2)\frac{n}{m}] \ge 1 - \frac{2}{c^2}$. This completes the proof.
\end{proof}

\section{Las Vegas Algorithms for $k$-SUM}

This section uses the hashing results to derive space-efficient Las Vegas
algorithms for $k$-SUM problems. Specifically, we demonstrate how to reduce the
space usage of $k$-SUM algorithms using Corollary~\ref{corollary-univ}. We use
that result to derive a family of linear-space Las Vegas algorithms for $k$-SUM.
We then reapply that result to derive a set of sublinear-space Las Vegas
algorithms that we will use later to establish new time-space upper bounds for
SUBSET-SUM algorithms.

\begin{theorem}
\label{las-vegas-size-reduction}
  Let $A$ be a Las Vegas algorithm that solves $k$-SUM ($k \ge 3$) on $n$
  numbers in $T(n)$ time and $S(n)$ space where $T(n), S(n) \in poly(n)$, and
  let $\delta \le 1$ be an arbitrary constant. Then there is a Las Vegas
  algorithm $A'$ that solves $k$-SUM on $n$ numbers in
  $O(n^{k-\delta(k-1)}+n^{k-\delta(k-1)-1}T(n^{\delta}))$ time and
  $O(n^{\delta} + S(n^{\delta}))$ space.
\end{theorem}

This theorem allows us to reduce the space usage of a $k$-SUM algorithm by a
factor of $\delta$ at the cost of shrinking the gap between the running time and
$O(n^k)$.

\begin{proof}
  The key idea is that to use hashing to reduce the size of each set by a square
  root factor at each step. However, storing any of the intermediate sets of
  this computation defeats the purpose of hashing any further. To avoid this,
  we first determine all hash functions and values to shrink each set to the
  desired size, and then compute the final sets in one step.

  $A'$ will recursively construct a list $L$ whose elements are of the form \\
  $(h, v_1, v_2, \ldots, v_k)$, i.e. a hash function followed by $k$ hash
  values (one for each $S_i$). At any step, define the active set of $S_i$ to
  be \\
  $\tilde{S}_i = \{s \in S_i \mid h(s) = v_i \forall
            (h, v_1, v_2, \ldots, v_k) \in L\}$.
  Each element appended to $L$ reduces the size of all active sets, so elements
  can be repeatedly appended until the active sets are only $O(n^{\delta})$
  in size, at which point it is safe to invoke $A$. To handle the possibility
  that $\delta$ is not a perfect power of $\frac{1}{2}$, define the function
  $s(x) := max((\frac{1}{2})^x,\delta)$. Step $i$ of the algorithm will reduce
  the size of all active sets from $O(n^{s(i)})$ to $O(n^{s(i+1)})$.

  The recursive helper function $R$ will construct $L$ and then invoke $A$. It
  has access to all sets $S_i$ and does the following given a partially
  constructed $L$:
  \begin{enumerate}
    \item Let $\ell := |L|$. If $s(\ell) = \delta$, then compute
          $\tilde{S}_1, \tilde{S}_2, \ldots, \tilde{S}_k$ and call $A$ on them.
          Otherwise, the active sets
          $\tilde{S}_1, \tilde{S}_2, \ldots, \tilde{S}_k$, are guaranteed to
          each contain at most $(k+2)^2 n^{s(\ell)}$ elements.
    \item Let $V_\ell := (k+2)n^{s(\ell)-s(\ell+1)}$. Pick a random hash
          function $h \in H_1$ that maps to $V_\ell$ values. For each $S_i$ and
          possible hash value $v \in [V_\ell]$, iterate through all elements of
          $S_i$, consider only the ones in $\tilde{S}_i$, and count how many
          hash to the current $v$. If any count exceeds $(k+2)^2 n^{s(\ell+1)}$
          elements, pick another hash and try again.
    \item For each $(v_1, v_2, \ldots, v_{k-1}) \in [V_\ell]^{k-1}$ and $j=0, 1,
          \ldots k-1$, let $v_k$ equal $h(t)$, less the sum of all already
          selected $v_i$'s, less $j$ (mod $V_\ell$). Call $R$ on $L$ appended
          with $(h, v_1, v_2, \ldots, v_k)$.
  \end{enumerate}
  Algorithm $A'$ calls $R$ with $L = \emptyset$.

  \begin{correctness}
    We first prove the size guarantee made when calling $R$. $A'$ initially
    calls $R$ with $\ell = 0$ and sets of size $n \le (k+2)^2 n^{s(0)}$. $R$
    ensures that the hash it has chosen creates buckets that are no larger than
    $(k+2)^2 n^{s(\ell+1)}$ in size, so it may safely append an additional
    element to $L$ before making a recursive call to itself.
    
    We also want to show that if a solution exists, we will find it. Due to
    the almost linearity property of $H_1$, we know that a call to $R$ where
    each element of the solution is in an active set will in turn make some
    recursive call where the elements are still in active sets. Since our first
    call to $R$ is made with an empty $L$ (and hence with all elements in active
    sets), we know that any elements of a solution will begin in active sets
    and hence will be found by the algorithm.
  \end{correctness}

  \begin{runningtime}
    Checking that the buckets of a randomly-selected hash function are not too
    large takes $O(n^{1+s(\ell)-s(\ell+1)})$ time since the algorithm needs to
    perform a linear scan for each hash value $v \in [V_\ell]$. We apply
    Corollary~\ref{corollary-univ} with $c=k$, so we know the chance of a hash
    failing over a specific $S_i$ is at most $\frac{2}{k^2}$; the chance of it
    failing over any $S_i$, by a union bound, is at most $\frac{2}{k}$. Since
    $k \ge 3$, the expected number of hashes the algorithm needs to pick and
    check is at most three. Hence our expected time checking for hashes during a
    single call to $R$, not including recursive subcalls, is
    $O(n^{1+s(\ell)-s(\ell+1)})$.

    There is a single call where $\ell = 0$. Each recursive level of $R$ makes \\
    $O(n^{(k-1)(s(\ell)-s(\ell+1))})$ calls to the level below it. Hence, there
    are $O(n^{(k-1)(1-s(\ell))})$ calls to $R$ for a given $\ell$ (all the terms
    cancel) The total expected time checking for hashes during all calls with a
    given $\ell$ is therefore
    $O(n^{(k-1)(1-s(\ell)) + (1+s(\ell)-s(\ell+1))})$.
    
    \begin{align*}
      (k-1)(1-s(\ell)) + (1+s(\ell)-s(\ell+1))
        &= k-s(\ell)(k-2)-s(\ell+1) \\
        &\le k-s(\ell+1)(k-1) \\
        &\le k-\delta(k-1) \\
    \end{align*}

    Hence, the total expected time checking for hashes during all calls with
    a given $\ell$ is also $O(n^{k-\delta(k-1)})$. Since the algorithm only
    searches for hash functions for $\ell \in \{0, 1, \ldots,
    \lceil \log_2 \frac{1}{\delta} \rceil - 1\}$, the total expected running
    time checking for hashes overall is $O(n^{k-\delta(k-1)})$.
    
    When $s(\ell) = \delta$, we need to compute all $\tilde{S}_i$. From our
    previously-derived formula, we know that there are only
    $O(n^{(k-1)(1-\delta)})$ calls where this occurs. Computing all
    $\tilde{S}_i$ only requires a linear scan of each $S_i$, so we can do this
    in time $O(n^{k-\delta(k-1)})$.

    Finally, we invoke $A$ $O(n^{(k-1)(1-\delta)})$ times on sets of size
    at most $(k+2)^2 n^{\delta}$, so in total we use
    $O(n^{(k-1)(1-\delta)}T(n^{\delta}))$ time making calls to $A$.

    The total time taken is hence
    $O(n^{k-\delta(k-1)}+n^{k-\delta(k-1)-1}T(n^{\delta}))$.
  \end{runningtime}

  \begin{memoryusage}
    Notice that $L$ contains at most $\lceil \log_2 \frac{1}{\delta} \rceil$
    elements of size $(k+1)$ each, so it takes $O(1)$ space. The space needed
    to check the selected hash is also $O(1)$, since we compute a count for
    only a single hash value at a time.

    Invoking $A$ on sets of size at most $(k+2)^2 n^{\delta}$ requires
    only $O(n^{\delta}+S(n^{\delta}))$ space (to store the inputs along with
    the space needed by $A$).
  \end{memoryusage}

  This completes the proof.
\end{proof}

This family of hash functions does particularly well when applied to 3-SUM. When
applied to the basic $O(n^2)$ time, $O(n)$ algorithm, the space-usage decreases
without any running-time cost:

\begin{reptheorem}{3-SUM-better}
  3-SUM on $n$ numbers can be solved by a Las Vegas algorithm in time $O(n^2)$
  and space $O(\sqrt{n})$.
\end{reptheorem}

\begin{proof}
  From Theorem~\ref{basic-3-sum}, we know 3-SUM can be solved in $T(n)=O(n^2)$
  time and $S(n)=O(n)$ space. We apply Theorem~\ref{las-vegas-size-reduction}
  with $\delta = 0.5$, which yields a Las Vegas algorithm that solves
  3-SUM in $O(n^{3-0.5(2)}+n^{3-0.5(2)-1}n^{0.5 \cdot 2})$, or $O(n^2)$ time and
  $O(\sqrt{n})$ space.
\end{proof}

Theorem~\ref{las-vegas-size-reduction} also yields a family of linear-space Las
Vegas algorithms for $k$-SUM problems, via the following intermediate corollary:

\begin{corollary}
\label{las-vegas-join}
  Let $A$ be a Las Vegas algorithm that solves $k_1$-SUM \\($k_1 \ge 3$) on $n$
  numbers in $\tilde{O}(n^{\alpha})$ time and $O(n)$ space for some constant
  $\alpha$.  Then there is a Las Vegas algorithm $A'$ that solves
  $(k_1 \cdot k_2)$-SUM on $n$ numbers in
  $\tilde{O}(n^{k_1k_2-k_1+1} + n^{k_1k_2-k_1+1+(\alpha-k_2)})$ time and
  $O(n)$ space.
\end{corollary}

\begin{proof}
  Apply Theorem~\ref{las-vegas-size-reduction} to $A$, choosing $\delta =
  \frac{1}{k_2}$. Hence, there is a Las Vegas algorithm $A''$ that solves
  $k_1$-SUM on $n$ numbers in \\
  $\tilde{O}(n^{k_1-(k_1-1)/k_2} +
             n^{k_1-(k_1-1)/k_2+\alpha/k_2-1})$ time and
  $O(n^{1/k_2})$ space.

  However, a $(k_1 \cdot k_2)$-SUM problem on $n$ numbers can be converted to a \\
  $k_1$-SUM problem on $n^{k_2}$ numbers. For $i \in \{1,2,\ldots,k_1\}$, we let
  $S'_i = \sum_{j=1}^{k_2} S_{(i-1)k_2+j}$ (the Minkowski sum of a block of
  $k_2$ sets), and we run $A''$ on the sets $S'_i$ with the same target $t$.
  Note that we do not actually store all elements of the sets $S'_i$, but rather
  compute them on demand in constant time as $A''$ requires them, in order to
  avoid using too much memory.

  Our algorithm $A'$ is to call $A''$ on the sets $S'_i$. Since these are
  $n^{k_2}$ in size, the algorithm $A'$ takes
  $\tilde{O}(n^{k_1k_2-k_1+1}+n^{k_1k_2-k_1+1+(\alpha-k_2)})$ time and $O(n)$
  space, as desired.

  This completes the proof.
\end{proof}

Corollary~\ref{las-vegas-join} can be used to find a linear-space algorithm for
$k$, given that it factors into $k_1$ and $k_2$ and that we already have an
algorithm for $k_1$-SUM that runs in linear space. If $k$ does not factor
nicely, it is possible to brute-force over one set to reduce to $(k-1)$-SUM:

\begin{lemma}
\label{las-vegas-induction}
  Let $A$ be an algorithm that solves $k$-SUM ($k \ge 3$) on $n$ numbers in
  $\tilde{O}(n^{\alpha})$ time and $O(n)$ space. Then there is an algorithm
  $A'$ that solves $(k+1)$-SUM on $n$ numbers in $\tilde{O}(n^{\alpha + 1})$
  time and $O(n)$ space.
\end{lemma}

\begin{proof}
  The algorithm $A'$ is to guess one element $s \in S_{k+1}$ of the solution and
  then to run $A$ on $S_1, \ldots, S_k$ for the remaining elements, which now
  need to sum to $t-s$.
\end{proof}

We now construct a function $f(k)$ such that we can produce a Las Vegas
algorithm that can solve $k$-SUM on $n$ numbers in $\tilde{O}(n^{f(k)})$ time and
$O(n)$ space.

\begin{definition}
  Let $f: \mathbb{Z}^+ \to \mathbb{Z}^+$. Let $f(1)=1$,
  $f(2)=1$, $f(3)=2$, and $f(4)=2$. For $k > 4$, let:
  \[ f(k) = \min_{\substack{
      k_1, k_2 \\
      k_1 \cdot k_2 = k
    }}
    \begin{cases}
      k_1k_2 - k_1 - k_2 + 1 + \max(f(k_1),k_2) \\
      f(k-1)+1
    \end{cases}
  \]
\end{definition}

\begin{corollary}
\label{las-vegas-linear-family}
  $k$-SUM on $n$ numbers can be solved in $\tilde{O}(n^{f(k)})$ time and
  $O(n)$ space by a Las Vegas algorithm.
\end{corollary}

\begin{proof}
  The base cases are covered by Theorem~\ref{basic-2-sum},
  Theorem~\ref{basic-3-sum}, and Theorem~\ref{basic-4-sum} (and $k=1$ is trivial).
  For all other $k$, we either get an algorithm from
  Corollary~\ref{las-vegas-join} or Lemma~\ref{las-vegas-induction}.
\end{proof}

\begin{table}[ht]
\caption{Time Complexity Upper Bounds for Linear-Space $k$-SUM Algorithms}
\centering
\begin{tabular}{| c | c | c |}
\hline
$k$ & $f(k)$ & $k-f(k)$\\ [0.5ex]
\hline
2 & 1 & 1\\
3 & 2 & 1\\ \cdashline{3-3}
4 & 2 & 2\\
5 & 3 & 2\\
6 & 4 & 2\\
7 & 5 & 2\\ \cdashline{3-3}
8 & 5 & 3\\
9 & 6 & 3\\
10 & 7 & 3\\
11 & 8 & 3\\
12 & 9 & 3\\
13 & 10 & 3\\
14 & 11 & 3\\ \cdashline{3-3}
15 & 11 & 4\\
\hline
\end{tabular}
\label{table:f(k)}
\end{table}

Table~\ref{table:f(k)} shows the first few values of $f(k)$. The difference
between $k$ and $f(k)$ is important because higher differences will permit
better time/space tradeoffs. Due to the construction of $f(k)$, this value
$k-f(k)$ is nondecreasing in $k$ (it is a maximum of its previous value and the
result of applying Corollary~\ref{las-vegas-join}). The values of $k$ where
$k-f(k)$ first increases to a new value $v$ ($k=8,15,24,32,40,54,\ldots$) occur
when $k$ factors evenly into $(v+1) \cdot f(v+1)$ (e.g. $15=5 \cdot 3$) or when
$k$ factors evenly into $(v+2) \cdot (f(v+2)-1)$ (e.g. $32 = 8 \cdot 4$),
whichever is smaller (the latter case occurs if $f(v+2) = f(v+1)$).

$f(x)$ has the following (coarse) upper bound:

\begin{lemma}
\label{f-bound}
  For all $x \ge 2$, $f(x) \le x - \sqrt{x} + 1$.
\end{lemma}

\begin{proof}
  Notice from Table~\ref{table:f(k)} that this is true for $x \in [2,8]$. We
  will now prove it for $x \ge 9$.

  Let $y^2$ be the largest perfect square that is at most $x$. By the definition
  of $f$, we know that $f(x) \le f(y^2 - 1) + (x - y^2 + 1)$. Hence, it suffices
  to show that $f(y^2 - 1) - y^2 \le -\sqrt{x}$.

  Since $x \ge 9$, $y+1 \ge 4$. Since $k-f(k)$ is nondecreasing in $k$,
  $(y+1)-f(y+1) \ge 2$. Simplifying yields $(y-1) \ge f(y+1)$.

  Notice that $y^2-1$ factors into $(y+1) \cdot (y-1)$. By our definition of
  $f$:

  \begin{eqnarray*}
    f(y^2-1) &\le& (y^2-1) - (y+1) - (y-1) + 1 + \max(f(y+1),y-1) \\
             &\le& y^2 - y - 1
  \end{eqnarray*}

  By our choice of $y$, though, this implies that:

  \[f(y^2-1) - y^2 \le -y - 1 \le -\sqrt{x}\]
  
  This completes the proof.
\end{proof}

\begin{repcorollary}{sqrt-linear}
  $k$-SUM on $n$ numbers can be solved in $\tilde{O}(n^{k-\sqrt{k}+2})$ time and
  $O(n)$ space by a Las Vegas algorithm.
\end{repcorollary}

\begin{proof}
  This is a direct consequence of Lemma~\ref{f-bound} combined with
  Corollary~\ref{las-vegas-linear-family}.
\end{proof}

Applying Corollary~\ref{las-vegas-size-reduction} once more to this linear-space
family yields sublinear algorithms:

\begin{reptheorem}{las-vegas-final}
  Let $\delta \le 1$. Then $k$-SUM on $n$ numbers can be solved in 
  $\tilde{O}(n^{k-\delta(k-1)} + n^{k-\delta(k-1) + (\delta f(k)-1)})$ time and
  $O(n^{\delta})$ space by a Las Vegas algorithm.
\end{reptheorem}

\begin{proof}
  Corollary~\ref{las-vegas-linear-family} states that there is a Las Vegas
  algorithm for $k$-SUM that runs in $\tilde{O}(n^{f(k)})$ time and $O(n)$
  space. Applying Corollary~\ref{las-vegas-size-reduction} then yields the
  desired result.
\end{proof}

\section{SUBSET-SUM Time-Space Tradeoffs}

Schroeppel and Shamir\cite{Schroeppel:1979:TTT:1382433.1382632} provided the
following reduction from SUBSET-SUM to $k$-SUM:

\begin{theorem}
\label{subset-sum-reduction}
  Let $A$ be an algorithm that solves $k$-SUM on $n$ numbers in
  $\tilde{O}(n^{\alpha k})$ time and $\tilde{O}(n^{\beta k})$ space for some
  constants $\alpha$ and $\beta$. Then SUBSET-SUM on $n$ numbers can be solved
  in $O^*(2^{\alpha n})$ time and $O^*(2^{\beta n})$ space.
\end{theorem}

\begin{proof}
  Consider the following algorithm $A'$:
  \begin{enumerate}
    \item Given a set $S$ with $n$ elements, divide it into $k$ sets $S_1, S_2,
          \ldots, S_k$ of $\frac{n}{k}$ elements each. For each set $S_i$,
          compute the set $T_i := \{\sum_{s \in S'_i} s \mid S'_i \subseteq
          S_i\}$. Run $A$ on $T_1, T_2, \ldots, T_k, t$.
  \end{enumerate}

  \begin{correctness}
    If there is some solution, the sum of its elements in any $S_i$ will wind
    up in some $T_i$, and hence $A$ will be able to find a solution that sums
    to $t$. Note that it is possible to backtrack and recover the original
    elements used to generate the elements of the $k$-SUM solution.
  \end{correctness}

  \begin{runningtime}
    We call $A$ on sets of size at most $2^{\frac{n}{k}}$, so $A'$ takes
    $O^*(2^{\alpha n})$ time.
  \end{runningtime}

  \begin{memoryusage}
    We call $A$ on sets of size at most $2^{\frac{n}{k}}$, so $A'$ takes
    $O^*(n^{\beta n})$ space.
  \end{memoryusage}

  This completes the proof.
\end{proof}

They also proved a theorem regarding SUBSET-SUM (as well as other problems
in a specific class of NP-hard problems) that allowed trading increased running
time in return for reduced space. Here is a $k$-SUM analogue of that result,
which allows further improvement our space-time upper bound on $k$-SUM (and
via Theorem~\ref{subset-sum-reduction}, SUBSET-SUM as well):

\begin{theorem}
\label{k-sum-time-space-tradeoff}
  Let $A$ be an algorithm that solves $k$-SUM on $n$ numbers in
  $T = \tilde{O}(n^{\alpha k})$ time and $S = \tilde{O}(n^{\beta k})$ space for some
  constants $\alpha$ and $\beta$. Then $k$-SUM on $n$ numbers can be solved in
  any time/space combination along the tradeoff curve
  $T' \cdot S'^{\frac{1 - \alpha}{\beta}} = \tilde{O}(n^k)$,
  $\Omega(n^{\alpha k}) \le T' \le \tilde{O}(n^k)$.
\end{theorem}

\begin{proof}
  Consider the following algorithm $A_\gamma$ ($0 \le \gamma \le 1$):
  \begin{enumerate}
    \item Divide each $S_i$ into $n^{1-\gamma}$ regions of consecutive elements
          of size $n^{\gamma}$.
    \item For each way to choose exactly one region from each $S_i$, run $A$ on
          that choice.
  \end{enumerate}

  In particular, when $\gamma = 0$, $A_\gamma$ is just a brute-force search,
  while when $\gamma = 1$, $A_\gamma$ reduces to algorithm $A$.

  \begin{correctness}
    We exhaustively search every combination of regions, and we know each
    element in a solution must appear in some region.
  \end{correctness}

  \begin{runningtime}
    Algorithm $A$ is called $n^{k(1-\gamma)}$ times on problems of size
    $O(n^{\gamma})$, so $A_\gamma$ uses
    $\tilde{O}(n^{k(1-\gamma) + \alpha \gamma k})$ time.
  \end{runningtime}

  \begin{memoryusage}
    Algorithm $A$ is called on problems of size $O(n^{\gamma})$, so $A_\gamma$
    uses $\tilde{O}(n^{\beta \gamma k})$ space.
  \end{memoryusage}

  Notice that:
  \begin{align*}
    T' \cdot S'^{\frac{1 - \alpha}{\beta}}
      &= \tilde{O}(n^{k(1-\gamma) + \alpha \gamma k} n^{\beta \gamma k \frac{1-\alpha}{\beta}}) \\
      &= \tilde{O}(n^{k - \gamma k + \alpha \gamma k + \gamma k - \alpha \gamma k}) \\
      &= \tilde{O}(n^k).
  \end{align*}

  This completes the proof.
\end{proof}

We have a family of space-efficient Las Vegas algorithms for $k$-SUM from
Corollary~\ref{las-vegas-final}. Applying
Theorem~\ref{k-sum-time-space-tradeoff} followed by
Theorem~\ref{subset-sum-reduction} yields a piecewise upper-bound curve for
SUBSET-SUM algorithms (due to the fact that $k$ must be integer). To better
understand the behavior of this curve, we formulate it as a tradeoff between
the exponents of $T$ and $S$, as follows:

\begin{reptheorem}{approximation-curve}
  There is a Las Vegas algorithm for SUBSET-SUM on $n$ numbers that runs in
  $O^*(2^{(1-\sqrt{\frac{8}{9}\beta})n})$ time and $O^*(2^{\beta n})$
  space, for $\beta \in [0, \frac{9}{32}]$.
\end{reptheorem}

We first prove a lemma:

\begin{lemma}
\label{las-vegas-approx}
  Given a constant $\beta \in [0, \frac{1}{4 \gamma^2}]$, there exists a $k$
  such that $k$-SUM on $n$ numbers is solved by a Las Vegas algorithm that runs
  in $T = \tilde{O}(n^{(1-\gamma \sqrt{\beta})k})$ time and
  $S = \tilde{O}(n^{\beta k})$ space, where
  $\gamma = \sqrt{\frac{8}{9}} \approx 0.942809$.
\end{lemma}

\begin{proof}
  Notice that $T \cdot S^{\frac{\gamma}{\sqrt{\beta}}} = \tilde{O}(n^k)$, so by
  Theorem~\ref{k-sum-time-space-tradeoff}, it suffices to show that there is
  a Las Vegas algorithm for $k$-SUM on $n$ numbers that runs in time $T'$ and
  space $S'$ where $T' \le T$ and $T' \cdot S'^{\frac{\gamma}{\sqrt{\beta}}} =
  \tilde{O}(n^k)$. For the remainder of the proof, we will let $\omega$ denote
  $\frac{\gamma}{\sqrt{\beta}}$.

  If $\omega \le 2$ then we are already done, since by Theorem~\ref{basic-4-sum}
  we have a solution to 4-SUM on $n$ numbers with $T' = \tilde{O}(n^2)$ and $S' =
  \tilde{O}(n)$ and our range for $\beta$ implies that $\gamma \sqrt{\beta}
  \le 0.5$. Hence we may assume that $\omega > 2$ for the remainder of the
  proof.

  By Corollary~\ref{las-vegas-final}, $k$-SUM on $n$ numbers can be solved in
  $T'' = \tilde{O}(n^{k-\delta(k-1)} + n^{k-\delta(k-1-f(k))-1})$ time and
  $S'' = \tilde{O}(n^{\delta})$ space by a Las Vegas algorithm. Choose
  $k = \lceil \omega \rceil + 1$ and $\delta = \frac{1}{\omega - 1}$. Since
  $\omega > 2$, we know that $k \ge 4$ and so $k-1-f(k) \ge 1$. Hence the
  running time $T''$ is in $\tilde{O}(n^{k-\delta-1})$.

  $\gamma$ should be chosen to guarantee that $\gamma \sqrt{\beta} k
  \le \delta+1$. Equivalently, $\gamma$ should be chosen such that:

  \begin{align*}
    \gamma \sqrt{\beta} k & \le \delta+1 \\
               \gamma^2 k & \le \omega(\delta+1) \\
               \gamma^2 k & \le \omega \frac{\omega}{\omega-1} \\
                 \gamma^2 & \le \frac{\omega^2}{(\omega-1)k} \\
                 \gamma^2 & \le \frac{\omega^2}{(\omega-1)(\omega+2)} \\
  \end{align*}

  The right-hand side is minimized when $\frac{(\omega-1)(\omega+2)}{\omega^2}$
  is maximized. Taking the derivative shows that this occurs when $\omega=4$,
  so it is safe to pick $\gamma = \sqrt{\frac{8}{9}}$.

  This completes the proof.
\end{proof}

Applying Theorem~\ref{subset-sum-reduction} to Lemma~\ref{las-vegas-approx}
directly yields Theorem~\ref{approximation-curve}.

The following graph demonstrates the previous best-known trade-off curve, found
by Schroeppel and Shamir \cite{Schroeppel:1979:TTT:1382433.1382632} (labeled as
basic 4-SUM) along with the piecewise upper-bound obtainable from
Corollary~\ref{las-vegas-final}. The graph also includes the time-space tradeoff
as given by Theorem~\ref{approximation-curve}.

\includegraphics[scale=0.65]{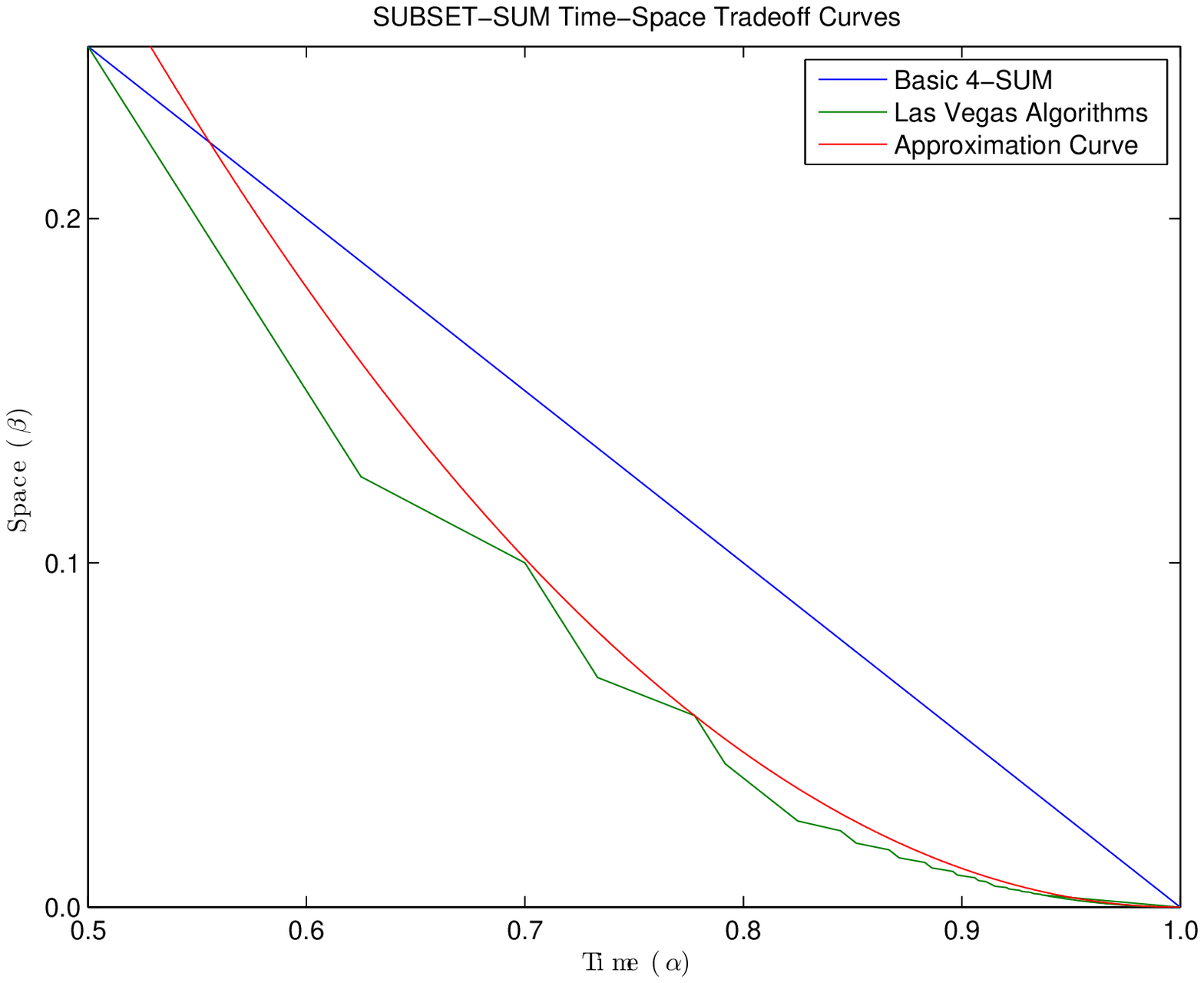}

\section{Conclusion}

An interesting open problem is whether there exists a deterministic algorithm
that runs in the same time for the $k$-SUM problem on $n$ numbers. It might be
easier to consider the $k$-XOR problem, which is identical except that the
elements are vectors from $\mathbb{F}_2^n$ instead of integers. For this
variant there is a simple linear universal family of hash functions, and so it
seems possible that there might be a way to derandomize the hash selection
process.

Another interesting question is whether the function $f(k)$ can be further
improved. The fact that there exists a $\tilde{O}(n^2)$ time and $O(n)$ space
algorithm for 4-SUM does not match the pattern found in the rest of the table,
suggesting that there might be more efficient algorithms for other $k$ as well.

\section*{Acknowledgements}

I would like to thank Ryan Williams for providing helpful pointers to existing
literature, insightful discussions, and proofreading.

\bibliographystyle{plain}
\bibliography{RANDOMIZED}

\begin{thebibliography}{1}

\bibitem{Baran:2005:SA:2145115.2145165}
Ilya Baran, Erik~D. Demaine, and Mihai P\v{a}tra\c{s}cu.
\newblock Subquadratic algorithms for {3SUM}.
\newblock In {\em Proceedings of the 9th International Conference on Algorithms
  and Data Structures}, WADS'05, pages 409--421, Berlin, Heidelberg, 2005.
  Springer-Verlag.

\bibitem{Dietzfelbinger:1996:UHK:646511.695324}
Martin Dietzfelbinger.
\newblock Universal hashing and k-wise independent random variables via integer
  arithmetic without primes.
\newblock In {\em Proceedings of the 13th Annual Symposium on Theoretical
  Aspects of Computer Science}, STACS '96, pages 569--580, London, UK, UK,
  1996. Springer-Verlag.

\bibitem{Gajentaan:2012:COP:2109239.2109580}
Anka Gajentaan and Mark~H. Overmars.
\newblock On a class of {$O(n^2)$} problems in computational geometry.
\newblock {\em Comput. Geom. Theory Appl.}, 45(4):140--152, May 2012.

\bibitem{Patrascu:2010:PFS:1873601.1873687}
Mihai P\u{a}tra\c{s}cu and Ryan Williams.
\newblock On the possibility of faster sat algorithms.
\newblock In {\em Proceedings of the Twenty-First Annual ACM-SIAM Symposium on
  Discrete Algorithms}, SODA '10, pages 1065--1075, Philadelphia, PA, USA,
  2010. Society for Industrial and Applied Mathematics.

\bibitem{Schroeppel:1979:TTT:1382433.1382632}
Richard Schroeppel and Adi Shamir.
\newblock A {$T \cdot S^2 = 0(2^n)$} time/space tradeoff for certain
  {NP}-complete problems.
\newblock In {\em Proceedings of the 20th Annual Symposium on Foundations of
  Computer Science}, SFCS '79, pages 328--336, Washington, DC, USA, 1979. IEEE
  Computer Society.

\end{thebibliography}

\end{document}